# A Magic Ratio Rule for Beginners: a Chemist's Guide to Quantum Interference in Molecules

Colin J. Lambert,[b] and Shi-Xia Liu*[a]

[a] Department of Chemistry and Biochemistry, University of Bern, Freiestrasse 3, CH-3012 Bern, Switzerland.
Fax: +41-31-631-3995    E-mail: liu@dcb.unibe.ch
[b] Quantum Technology Centre, Physics Department, Lancaster University, Lancaster LA1 4YB, UK
E-mail: c.lambert@lancaster.ac.uk
Prof. Colin J. Lambert and Dr. Shi-Xia Liu

**Abstract**
This overview will give a glimpse into chemical design principles for exploiting quantum interference (QI) effects in molecular-scale devices. Direct observation of room temperature QI in single-molecule junctions has stimulated growing interest in fabrication of tailor-made molecular electronic devices. Herein, we outline a new conceptual advance in the scientific understanding and technological know-how necessary to control QI effects in single molecules by chemical modification. We start by discussing QI from a chemical viewpoint and then describe a new magic ratio rule (MRR), which captures a minimal description of connectivity-driven charge transport and provides a useful starting point for chemists to design appropriate molecules for molecular electronics with desired functions. The MRR predicts conductance ratios, which are solely determined by QI within the core of polycyclic aromatic hydrocarbons (PAHs). The manifestations of QI and related quantum circuit rules for materials discovery are direct consequences of the key concepts of weak coupling, locality, connectivity, mid-gap transport and phase coherence in single-molecule junctions.

**Introduction**

How does electricity flow through molecules? To answer this question experimentally, strategies are need for contacting single molecules to source and drain electrodes separated by a nanogap and then passing a current from the source to the drain *via* the molecule.[1] In the literature, nanogaps have been realised using scanning-tunnelling-microscopy-based break junctions, conducting probe atomic force microscopy break junctions and mechanically controllable break junctions.[1g, 1h, 2] Fundamental studies of charge transport properties of single molecules trapped between two metallic electrodes have demonstrated clear correlations between molecular structure and function.[3] Scalability remains one of great challenges, which is of high importance for the fabrication of practical molecular electronic devices. Consequently, silicon-based platforms have been developed,[4] but so far such technologies remain in their infancy. To circumvent some limitations of metallic and silicon-based electrodes, single molecule junctions have been fabricated between carbon-based electrodes such as carbon nanotubes[5] and graphene.[6] In particular, graphene electroburnt junctions have been shown to deliver stable electrode gaps below 5 nm,[7] which allow electrostatic gating through buried or side gates. For the purpose of attaching molecules to different types of electrodes, the ends of the molecule should be terminated by 'anchor groups', which bind opposite ends of the molecule to the source and drain electrodes. Over the last two decades, a variety of anchor groups have been explored.[1a, 1e, 8] For instance, thiols,[9] pyridines,[10] amines,[11] methyl sulfides[12] and direct gold-carbon bonds[13] have been utilised in metal-molecule-metal junctions. Amine-terminated molecules can bridge nanogaps between carboxylic acid-functionalized carbon nanotubes[5b] while aromatic planar anchor groups including anthracene [7a] and pyrene[14] are of interest due to their binding ability to graphene electrodes *via* π-π stacking and van der Waals interactions. Having established stable anchors to the electrodes, the passage of electricity through single molecules requires making choices for the remainder of the molecule, which typically involves a central aromatic functional subunit attached to the anchor groups *via* spacers. Clearly, many different factors including anchor groups, molecular lengths, the nature of spacers and electronic structures of the aromatic subunits, affect drastically the charge transport properties of molecular devices. In most cases, charge transport measurements are interpreted in terms of an off-resonant tunneling process that disregards quantum interference (QI) effects.

During the past few years, a large body of evidence has accumulated, which demonstrates that the passage of electricity through molecules is controlled by QI, even at room temperature,[15] and furthermore, QI can be exploited to control and enhance electrical and thermoelectrical properties of single molecules.[15d, 16] As a consequence, fine tuning of QI effects has attracted a great deal of interest, both theoretically and experimentally. Despite efforts to control the flow of charge at the quantum level by modifying electronic structure and molecular topology through organic synthesis, a direct correlation between QI and structure remains ambiguous. It is both timely and desirable to develop qualitative design rules for tailor-made single-molecule devices. Recently we and our colleagues developed a new "magic ratio rule" (MRR), which captures the role of connectivity in determining the charge transport properties of polycyclic aromatic hydrocarbons (PAHs) as well as their heteroatom-substituted systems.[15b, 17] Evidence for QI is mostly indirect and obtained by comparing transport properties of homologous series of related molecules,[15f-h] rather than using external electric, magnetic or mechanical gating to control QI within a single molecule. In other examples, direct evidence of room temperature QI is obtained by manipulating the charge state of a molecule in three-terminal devices or in electrochemical environments.[15j] Before discussing these examples in more detail, we start by describing some basic concepts of QI, which are relevant to the flow of electricity and heat. We present these concepts using a language more familiar to chemists than to physicists, emphasizing the difference between inter-orbital and intra-orbital QI. One aim of the following text is to introduce Green's functions for molecular-scale transport in an intuitive and non-mathematical manner.

Ultimately, QI effects are related to the shapes and energies of the molecular orbitals, and can therefore be manipulated by chemical design. However when many orbitals contribute, focusing on the contributions from individual orbitals becomes cumbersome and it is fruitful to introduce new 'magic ratio rules', which account for many orbital QI effects in a simple and intuitive manner.

## Green's functions for beginners

Of course in single molecules, QI is everywhere. Molecular orbitals (MOs) such as the highest occupied molecular orbital (HOMO) of pyrene (see Fig. 1) are obtained by solving the Schrödinger equation and are themselves a result of constructive QI. In what follows, we shall refer to this as "intra-orbital QI." Our task is to understand the relationship between these MOs and the flow of electricity, and to understand how they can be manipulated by chemical design to optimise transport properties. The isosurface in Fig. 1 shows that the HOMO has regions of positive amplitude (coloured red) and regions of negative sign (coloured blue). Furthermore it has regions of high amplitude such as the regions of space $r_5$ and $r_8$ near atoms 5 and 8, and regions of low amplitude such as the region of space $r_2$ near atom 2. Such MOs are analogous to the wave patterns in your coffee cup, which occur when the cup is placed on a vibrating table whilst travelling on a train, or a sound wave pattern formed by an echo in an empty auditorium, or the shape of water waves created in a wave tank.

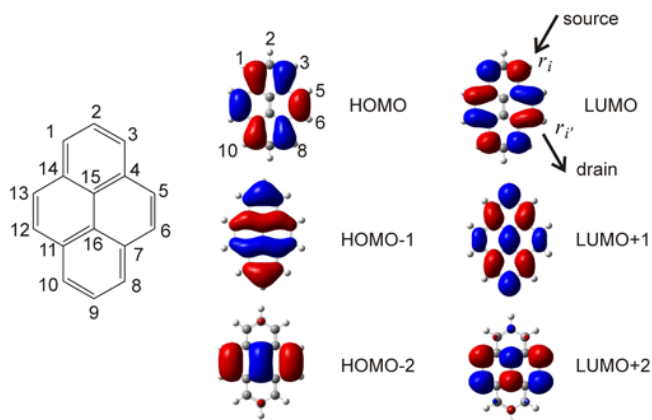

**Figure 1**. A lattice representation and frontier molecular orbitals of pyrene.

The basic ideas used to relate MOs of isolated molecules to the flow of electricity when located within a junction, were introduced in ref 1f, where the key concepts of
- weak coupling
- locality
- connectivity
- mid-gap transport
- phase coherence

were utilised.

To highlight the relevance of 'weak coupling,' we note that if an aromatic molecule such as pyrene is strongly coupled to two electrodes, then the identity of the molecule is lost and transport properties can only be computed by treating the molecule, electrodes and their couplings on an equal footing in a holistic manner. However if a central aromatic subunit such as pyrene is weakly coupled to the anchor groups *via* spacers such as acetylene, then the characteristics of the pyrene are preserved and the MOs of isolated pyrene can be used to predict transport properties. This 'weak coupling' concept is crucial to the discussion below.

The concept of 'locality' recognises that when a current flows through an aromatic subunit, the points of entry and exit are localised in space. For example in Fig. 1, the current enters in the region of space in the vicinity of position $r_i$ and exits from the region of space near $r_{i'}$.

The concept of 'connectivity' recognises that spacers can be attached to different parts of a central subunit with atomic accuracy and therefore it is of interest to examine how the flow of electricity depends on the choice of connectivity to the central subunit. For example if $\sigma_{ii'}$ is the electrical conductance of a single-molecule junction when the current enters and leaves a subunit at locations $r_i$ and $r_{i'}$, and $\sigma_{jj'}$ is the electrical conductance when the current enters and leaves at locations $r_j$ and $r_{j'}$, then it is of interest to examine conductance ratios such as $\sigma_{ii'}/\sigma_{jj'}$. Remarkably, if the coupling to the central subunit is weak, then such ratios are largely determined by MOs of the isolated subunit.

The concept of 'mid-gap transport' is an acceptance of the fact that unless a molecular junction is externally gated by an electrochemical environment or an electrostatic gate, then the energy $E$ of electrons flowing through the molecule is usually located in the vicinity of the centre of the HOMO-LUMO gap and therefore transport takes place in the co-tunnelling regime. In other words, transport is usually 'off-resonance'.

Finally, the concept of 'phase coherence' recognises that in this co-tunnelling regime, the phase of electrons is usually preserved as they pass through a molecule and therefore transport is controlled by superpositions of MOs of the type shown in Fig. 1. When these conditions are satisfied, electricity flow can be understood using the intuitive picture described below. Once the intuitive picture is established, one can systematically discuss what happens when each of the above conditions is relaxed.

To understand the relationship between MOs and electrical conductance, consider a fictitious subunit, which possesses only a single MO, which we denote $\psi^H(r)$, such as that of Fig. 1 and is weakly coupled to the anchors. Clearly the only QI in such a molecule is intra-orbital QI, because only one MO is present. If the current enters at the point $r_i$ and exits at the point $r_{i'}$, as shown in Fig. 1, then the electrical conductance $\sigma_{ii'}$ will be proportional to $(\psi^H(r_i)\psi^H(r_{i'}))^2$. This means that to obtain a large conductance, the amplitude of the MO should be large at both the entry and exit points. In the case of pyrene, the MO amplitude is small at position $r_2$ and therefore if either of the anchor groups is connected to this atom *via eg* a triple bond, the conductance will be low. On the other hand for entry and exit connectivities such as 5 and 8, the MO amplitude is large at both $r_5$ and $r_8$, so this connectivity corresponds to high electrical conductance. In summary, for this single-orbital molecule, the electrical conductance $\sigma_{ii'}$ is proportional to $(g_{ii'})^2$, where

$$g_{ii'} = c_H \psi^H(r_i) \psi^H(r_{i'}) \quad (1)$$

and $C_H$ is a constant of proportionality. Consequently, the ratio of two conductances corresponding to different connectivities is given by

$$\frac{\sigma_{ii'}}{\sigma_{jj'}} = \left(\frac{g_{ii'}}{g_{jj'}}\right)^2 \quad (2)$$

Obviously, the conductance ratio in this single-orbital example is independent of the choice of $C_H$. On the other hand, with an appropriate choice of $C_H$, the quantity $g_{ii'}$ is a Green's function of our fictitious single-orbital molecule. If the energy of the MO $\psi^H(r)$ is $E_H$ and the energy of electrons flowing through the molecule is $E_F$ then the appropriate choice is

$$C_H = \frac{1}{E_F - E_H} \quad (3)$$

In practice, the energy $E_F$ usually coincides with the Fermi energy of the external electrodes. As noted above, transport is usually off-resonance, which means that $E_F - E_H \neq 0$.

Although the form of $C_H$ given in equation (3) does not affect the right hand side of equation (2) for a single-MO subunit, it does become relevant when a subunit possesses more than one MO, because it controls the way in which MOs interfere with each other in a molecular junction. We refer to this type of QI as "inter-orbital QI". This new QI between MOs, which is absent in the isolated subunit, becomes relevant when the molecule is placed in a junction and electrons pass through the molecule from one connection point to another. To demonstrate how inter-orbital QI manifests itself in a molecular junction, consider a fictitious subunit possessing two MOs, such as the pyrene HOMO $\psi^H(r)$ of energy $E_H$ and LUMO $\psi^L(r)$ of energy $E_L$, shown in Fig 1. In this case, equation (2) is still satisfied, but equation (1) is replaced by

$$g_{ii'} = C_H \psi^H(r_i) \psi^H(r_{i'}) + C_L \psi^L(r_i) \psi^L(r_{i'}) \quad (4)$$

where $C_L = 1/(E_F - E_L)$. When combined with equation (2), this equation reveals that the flow of electrons through the two-MO subunit is controlled by inter-orbital QI arising from the superposition of MOs. Since MOs can be positive or negative at different locations, the two terms on the right hand side of equation (4) could have opposite signs (corresponding to destructive QI) or the same sign (corresponding to constructive QI). Since the amplitudes $C_H$ and $C_L$ depend on the energies of the MOs relative to the Fermi energy, this new form of QI can be manipulated and exploited by tuning these energies. Clearly if $E_F \approx E_L$, then transport is LUMO dominated, whereas if $E_F \approx E_H$, it is HOMO dominated. In these cases, intra-orbital QI is the most important. On the other hand for mid-gap transport, where $E_F = (E_H + E_L)/2$, $C_L = -C_H = 1/\delta$, where $\delta$ is half the HOMO-LUMO gap given by $\delta = (E_L - E_H)/2$. In this case, equation (4) yields for the mid-gap Green's function

$$g_{ii'} = C_L \left[ \psi^H(r_i)\psi^H(r_{i'}) - \psi^L(r_i)\psi^L(r_{i'}) \right] \quad (5)$$

and the constant $C_L$ cancels in the conductance ratio formula (2). This demonstrates that mid-gap conductance ratios are independent of the size of the HOMO-LUMO gap, even though the absolute values of the conductances are gap dependent.

Equation (5) reveals that for mid-gap transport, constructive interference occurs when the HOMO $\psi^H(r_i)\psi^H(r_{i'})$ and LUMO $\psi^L(r_i)\psi^L(r_{i'})$ products have opposite signs, whereas QI is destructive when they have the same sign. This allows us to spot connectivities with high or low conductances. For example, by inspection of Fig. 1, we can make the following observations:

1. Clearly, if current enters or leaves *via* the 'backbone sites' 2, 15, 16, 9 then the conductance will be low, because irrespective of their signs, intra-orbital QI causes the magnitudes of the HOMO and LUMO MOs on these sites to be small.
2. The HOMO product $\psi^H(r_1)\psi^H(r_8)$ is negative, whereas the LUMO product $\psi^L(r_1)\psi^L(r_8)$ is positive, so this connectivity corresponds to constructive inter-orbital QI and $\sigma_{1,8}$ will be high. The same conclusion applies to *eg* $\sigma_{5,6}$.
3. The HOMO product $\psi^H(r_{10})\psi^H(r_8)$ is negative and the LUMO product $\psi^L(r_{10})\psi^L(r_8)$ is also negative, so this connectivity corresponds to destructive inter-orbital QI and $\sigma_{10,8}$ will be low. The same conclusion applies to any odd-odd or even-even connectivity.

Equation (4) reproduces the main features found in many literature calculations of electrical conductance. To demonstrate this, it is useful to explore the energy dependence of electrical conductance by introducing the following short-hand notation for the MO products:

$$a_H = \psi^H(r_i)\psi^H(r_{i'}) \quad \text{and} \quad a_L = \psi^L(r_i)\psi^L(r_{i'}) \quad (6)$$

So that equation (4) can be written

$$g_{ii'}(E_F) = \frac{a_H}{E_F - E_H} + \frac{a_L}{E_F - E_L} \quad (7)$$

This expression shows how the physics of the electrodes, (contained in $E_F$) combines with the chemistry of the subunit (contained in the MOs) to control the flow of electricity through molecules. Clearly the constructive or destructive nature of the QI depends both on the relative signs of the MO products $a_H$ and $a_L$ and on the relative signs of the denominators $E_F$ - $E_H$ and $E_F$ - $E_L$. (Note that for the current description of transport within the gap, both of these denominators are non-zero.) With this caveat, equation (7) describes the main contribution to the flow of electricity from the non-degenerate HOMO and LUMO of any weakly-coupled subunit and therefore it is worth simplifying it by introducing the dimensionless energy

$$\varepsilon_F = \frac{E_F - \frac{(E_H + E_L)}{2}}{\delta} \qquad (8)$$

where $\delta = (E_L - E_H)/2$. Clearly $\varepsilon_F$ is simply the Fermi energy relative to the middle of the HOMO-LUMO gap, in units of half the HOMO-LUMO gap $\delta$ and is independent of connectivity. This notation allows equation (7) to be written

$$g_{ii'}(E_F) = \frac{1}{\delta} t_{ii'} \qquad (9)$$

where

$$t_{ii'} = \left[ \frac{a_H}{\varepsilon_F + 1} + \frac{a_L}{\varepsilon_F - 1} \right] \qquad (10)$$

Clearly $t_{ii'}$ depends on connectivity through the MO products $a_H$ and $a_L$. In what follows, we refer to $t_{ii'}$ as a core transmission amplitude and define the corresponding transmission coefficient $\tau_{ii'}$ by $\tau_{ii'} = |t_{ii'}|^2$. In terms of these dimensionless quantities, equation (2) becomes

$$\frac{\sigma_{ii'}}{\sigma_{jj'}} = \frac{\tau_{ii'}}{\tau_{jj'}} \qquad (11)$$

In summary, conductance ratios are obtained from a simple core transmission amplitude $t_{ii'}$, which captures many of the key features of quantum transport through molecules. This function involves the dimensionless energy $\varepsilon_F$ and dimensionless connectivity-dependent parameters $a_H$ and $a_L$.

From equation (10), we obtain the following rule: "Perfect destructive inter-orbital QI can only occur if MO products $a_H$ and $a_L$ have the same sign."

To demonstrate this, we note that from equation (10), $t_{ii'} = 0$, when $\varepsilon_F = \alpha_{ii'}$, where

$$\alpha_{ii'} = \frac{(a_H - a_L)}{(a_H + a_L)} \qquad (12)$$

which corresponds to perfect destructive interference. Since we are describing transport within the gap, where $\varepsilon_F$ is confined to the range, $-1 < \varepsilon_F < +1$ perfect destructive interference can only occur if $|\alpha| < 1$, which from equation (12) means that $a_H$ and $a_L$ have the same sign. Hence we conclude that there are only two qualitative different scenarios for transport through molecules when the Fermi energy lies in the HOMO-LUMO gap, corresponding to either $|\alpha| < 1$ or $|\alpha| > 1$, as shown in Fig 2.

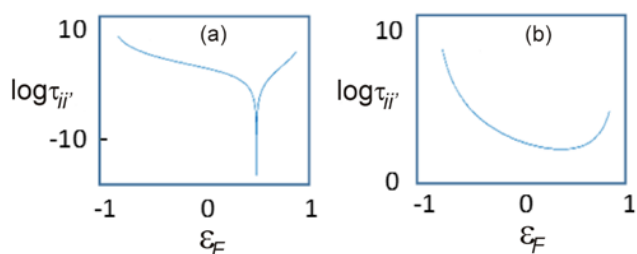

**Figure 2**. Plots (on a logarithmic scale) of $\tau_{ii'} = |t_{ii'}|^2$ (from equation 10) versus $\varepsilon_F$, for (a) $\alpha = 0.5$ and (b) $\alpha = 1.5$. The former shows destructive QI, while the latter does not.

These plots capture the generic features of many such curves found in the literature. They can be made to resemble literature results even more closely by recognising that coupling a molecular core to external electrodes introduces level broadening, whose generic effect can be captured by introducing a small positive imaginary term $i\eta$ into the denominators of equation (10), which now becomes

$$t_{ii'} = \left[\frac{a_H}{\varepsilon_F + 1 + i\eta} + \frac{a_L}{\varepsilon_F - 1 + i\eta}\right] \tag{13}$$

Finally, as shown in Fig. 1, molecules possess many orbitals (labelled n=0, +/-1, +/-2,…. *etc*, with energies $E_n$), beyond a simple HOMO (corresponding to n=0) and LUMO (corresponding to n=1). In this case equation (7) must be modified and takes the form

$$g_{ii'}(E_F) = \sum_n \frac{a_n}{E_F - E_n} \tag{14}$$

To estimate the contribution for these other MOs, consider the case where $E_F$ coincides with the centre of the HOMO-LUMO gap. In that case, if the level spacing is approximately constant, then $E_n = (2n-1)\delta$ and equation (14) becomes

$$g_{ii'}(E_F) = \frac{1}{\delta} \sum_n \frac{a_n}{(1 - 2n)} \tag{15}$$

As an example, for a molecule with four MOs, this becomes

$$g_{ii'}(E_F) = \frac{1}{\delta}\left[\frac{a_H}{(+1)} + \frac{a_L}{(-1)} + \frac{a_{-1}}{(+3)} + \frac{a_{+2}}{(-3)}\right]$$

Hence the contributions from the LUMO+1 and HOMO-1 are on the scale of one third of the contributions from the HOMO and LUMO and in general cannot be neglected. In practice, if the HOMO and LUMO exhibit perfect destructive interference at certain energy $\varepsilon_F = \alpha_{ii'}$ and then the contributions from other orbitals may either shift the energy at which destructive interference occurs or may eliminate the destructive QI completely.

**Magic ratio rule (MRR)**

When many orbitals contribute to inter-orbital QI, the right hand side of equation (15) becomes difficult to interpret and therefore it is useful to compute the Green's function $g_{ii'}(E_F)$ using an alternative (though mathematically-equivalent) approach based on tables of 'magic numbers,' whose validity again rests on the key concepts of weak coupling, locality, connectivity, mid-gap transport and phase coherence. For alternant polyaromatic hyrdrocarbons (PAHs) and related molecules obtained by heteroatom substitution, these tables predict ratios of statistically-most-probable conductances corresponding to different connectivities $r_i r_{i'}$ and $r_j r_{j'}$ provided the statistical properties of metal-molecule interfaces at single-molecule junctions are independent of connectivity. The connectivity-dependent core transmission $\tau_{ii'}$ can be calculated by introducing tables of 'magic integers' $M_{ii'}$, giving $\tau_{ii'} = (M_{ii'})^2$. When the Fermi energy of the electrodes lies close to the center of the HOMO−LUMO gap, the conductance ratio is equal to $(M_{ii'}/M_{jj'})^2$. We call this a "magic ratio rule" (MRR). For further details of how to construct M-Tables, such as that shown in figure 3, the reader is referred to our recent papers,[15b, 17] in which the accuracy of MRR for a range of PAHs as well as heteroatom-substituted systems has been verified experimentally.

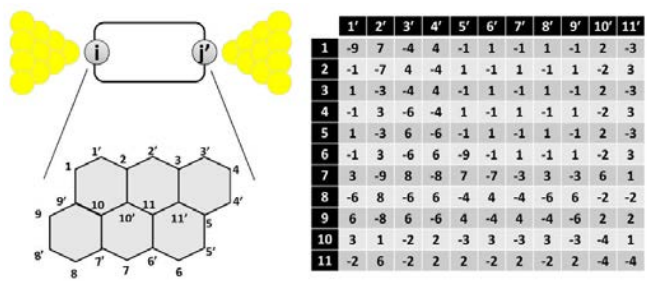

**Figure 3.** An example of the bipartite anthanthrene lattice together with its M-Table.

As an example, for the two anthanthrene-based molecules illustrated in Fig. 4, one electrode is connected to a site $r_i$ and the other is connected to a site $r_{i'}$ of the anthanthrene core, and so we assign the core a magic integer $M_{ii'}$. Two molecules with the same aromatic core but different pairs of electrode connection sites ($r_i r_{i'}$ and $r_j r_{j'}$, respectively) have different magic integers $M_{ii'}$ and $M_{jj'}$ as shown in the M-table.[17c] For instance, molecule **1** is connected through sites 1 and 5' and possesses a magic number -1, while molecule **2** is connected through sites 2' and 7 and possess a magic number -9 (Fig. 4). Consequently, the conductance of **2** is predicted to be a factor of $9^2/1^2 = 81$ higher than that of **1**. Mechanically controllable break junction (MCBJ) measurements (Fig. 4) reveal two distinct conductance values for **1** and **2** at $10^{-6.7\pm0.7}G_0$ and $10^{-4.8\pm0.6}G_0$, respectively, giving a conductance ratio of approximately 79, which matches well with the MRR. These results have further been rationalized by DFT calculations.[17c]

As a second example, the conductance ratio of two pyrene derivatives that are linked to Au electrodes through 1,8 and 2,9 sites, respectively, is predicted to be 9/1 by the MRR combined with the magic number table shown in Fig. 5. For

comparison, the experimental conductance ratio was measured to be 8/1 by using a MCBJ set-up.[17a]

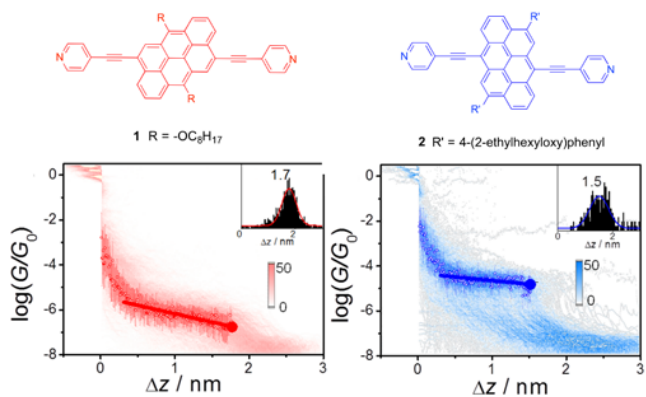

**Figure 4.** 2D conductance histograms and stretching distance distributions (inset) for **1** (red) and **2** (blue) using THF/mesitylene.

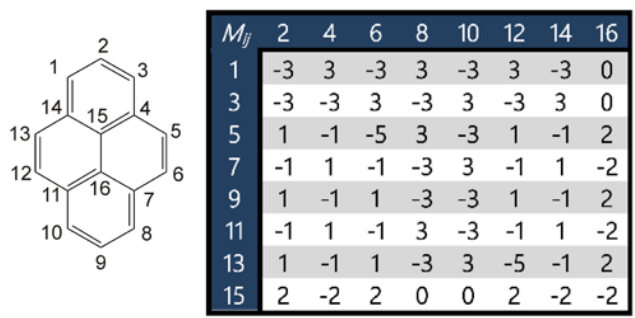

**Figure 5.** An example of the bipartite pyrene lattice together with its M-Table.

### Effect of heteroatom substitution on QI

Inspired by the success of the MRR for PAHs, we further verified and expanded this theory by studying the change in conductance when "parent" phenylene ethylene-type molecules (*meta*- and *para*-OPE) are modified to yield "daughter" molecules by inserting one nitrogen atom into the central benzene ring at different positions as shown in Fig. 6.[15b, 17b]

As depicted in Fig.7a and b, the measured conductance values are arranged in the following order: **M2 > M3 >M1 = *m*-OPE**, indicating that destructive interference in the meta-connected core of m-OPE can be alleviated to some extent, depending on the position of heteroatom substitution in the central phenyl core. In contrast, both ***P*** and ***p*-OPE** show a similar conductance, suggesting that constructive interference is negligibly affected by heteroatom substitution.

These experimental results can be verified by DFT calculations. As illustrated in Fig. 7c and d, near the gap centre ($E = 0$), there is a significant change in the core transmission coefficient for **M2** and **M3** compared to **M1** and *m*-**OPE**. The former appear non-zero, whereas the latter remain zero at $E = 0$. On the contrary, the core transmission coefficients of ***P*** and ***p*-OPE** are non-zero and overlap with each other. It can therefore be deduced that the destructive QI can be alleviated by the heteroatom substitution whereas constructive QI is almost unaffected. For **M1**-**M3**, the experimental conductance ratios compare well with core transmission ratios, and are equal to the square of the ratio of two "magic integers". Based on these results, the conductance ratios are determined by connectivity due to QI within the PAH core despite of the presence and absence of a heteroatom.

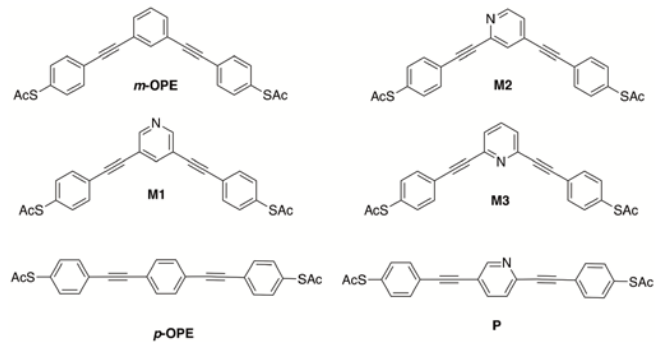

**Figure 6.** Chemical structures of the investigated molecules. The top left ***m*-OPE** and bottom left ***p*-OPE** are the parent molecules.

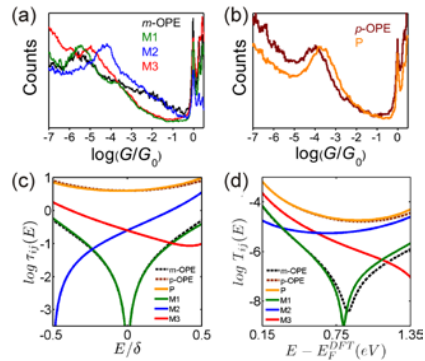

**Figure 7.** (a,b) All- data-point 1D conductance histograms constructed from 1000 MCBJ traces of each molecule; c) Core transmission coefficients $\tau_{ij}(E)$ of each molecule against $E/\delta$, where $\delta$ is half of the HOMO–LUMO gap of the parental core, using DFT transport approach implemented in Gollum;[18] d) The calculated transmission coefficients $\tau_{ij}(E)$ of each molecule, connected to gold electrodes using the mean-field Hamiltonian from Siesta.[19] Dashed lines correspond to "parents" and solid lines to "daughters".

## Conclusion

By invoking the key concepts of weak coupling, locality, connectivity, mid-gap transport and phase coherence, we have shown that inter-orbital QI can be understood by examining superpositions of MOs, as described by equation (10) and more generally equation (14). When many MOs are involved, the effects of QI are conveniently described using magic number tables, which reveal how single-molecule conductances depend on the connectivity to their cores. As illustrated by the almost two-orders-of-magnitude conductance ratio of molecules **1** and **2** of figure 4, these connectivity-dependencies are highly non-classical, since such a connectivity change would barely affect the conductance of a classical resistive network. The MRR is a simple, parameter-free, analytic theory, which captures the QI patterns within the hearts of PAH molecules at the mid-point of the HOMO−LUMO gap. It states that mid-gap conductance ratios of molecules equal to the square of the ratio of their magic numbers. This theory has been verified by comparison with measured conductance ratios of molecules with bipartite cores such as anthanthrene, anthracene, naphthalene and pyrene as well as non-bipartite cores such as azulene,[17a] and further generalized by heteroatom substitution of PAHs, where it has also been demonstrated that destructive QI can be alleviated by heteroatom substitution.[15b] These studies show that the MRR provides a useful starting point for chemists to design appropriate molecules for molecular electronics with desired functions and improved performance. As an example, from a measurement of the conductance of **1** in Fig. 4 and from knowledge of the magic numbers of **1** and **2**, the conductance of **2** could be predicted ahead of its synthesis. Finally, we note that weak coupling, locality, connectivity, mid-gap transport and phase coherence can be utilized to yield further quantum circuit rules, which relate the transport properties of molecules of the form A-B-C, B-A-C and B-C-A.[20] Such rules are also useful for the purpose of materials discovery, since from measurements of the electrical conductance or Seebeck coefficient of two such molecules, the electrical and thermoelectrical properties of a third can be predicted.

## Acknowledgements

This work was supported by the European Commission (EC) FP7 ITN "MOLESCO" (project no. 606728), UK EPSRC (grant nos. EP/M014452/1 and EP/N017188/1)

.

**Figure legends**

Figure 1. A lattice representation and frontier molecular orbitals of pyrene.

Figure 2. Plots (on a logarithmic scale) of $\tau_{ii'} = |t_{ii'}|^2$ (from equation 10) versus $\varepsilon_F$, for (a) $\alpha=0.5$ and (b) $\alpha=1.5$. The former shows destructive QI, while the latter does not.

Figure 3. An example of the bipartite anthanthrene lattice together with its M-Table.

Figure 4. 2D conductance histograms and stretching distance distributions (inset) for **1** (red) and **2** (blue) using THF/mesitylene.

Figure 5. An example of the bipartite pyrene lattice together with its M-Table.

Figure 6. Chemical structures of the investigated molecules. The top left m-OPE and bottom left p-OPE are the parent molecules.

Figure 7. (a,b) All- data-point 1D conductance histograms constructed from 1000 MCBJ traces of each molecule; c) Core transmission coefficients $\tau_{ij}(E)$ of each molecule against $E/\delta$, where $\delta$ is half of the HOMO–LUMO gap of the parental core, using DFT transport approach implemented in Gollum;[18] d) The calculated transmission coefficients $\tau_{ij}(E)$ of each molecule, connected to gold electrodes using the mean-field Hamiltonian from Siesta.[19] Dashed lines correspond to "parents" and solid lines to "daughters".

**Table of Contents**

By invoking the key concepts of weak coupling, locality, connectivity, mid-gap transport and phase coherence, we describe a magic ratio rule (MRR), which captures a minimal description of connectivity-driven charge transport and provides a useful starting point for chemists to design appropriate molecules for molecular electronics with desired functions.

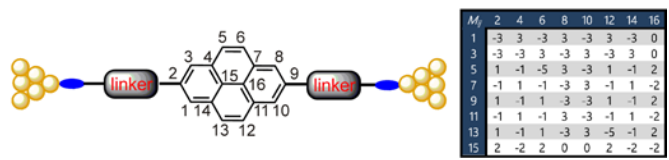

**Keywords:** molecular electronics • quantum interference • heteroatom effect • connectivity • single-molecule transport